\documentclass{article}
\usepackage{spconf,amsmath,graphicx}
\usepackage{booktabs}
\usepackage{pgfplots}
\usepackage{caption,subcaption}
\usepackage{color}
\usepackage{amssymb}
\usepackage{paralist}
\usepackage{cancel}
\usepackage{url}
\usepackage[inline]{enumitem}
\pgfplotsset{compat=1.18}
\usepackage{hyperref}
\usepackage{multirow}
\usepackage{soul}

\def\llmtospeech{\textit{LLM2Speech}}
\def\llmtophones{\textit{LLM2PnP}}
\def\phonetospeech{\textit{PnP2Speech}}

\DeclareMathOperator{\word}{word}

\title{SPEAK WHILE YOU THINK: \\ STREAMING SPEECH SYNTHESIS DURING TEXT GENERATION}

\name{Avihu Dekel, Slava Shechtman, Raul Fernandez, David Haws, Zvi Kons, Ron Hoory}
\address{IBM Research}
\begin{document}
%
\maketitle
\begin{abstract}
Large Language Models (LLMs) demonstrate impressive capabilities, yet interaction with these models is mostly facilitated through text.
Using Text-To-Speech to synthesize LLM outputs typically results in notable latency, which is impractical for fluent voice conversations.
We propose \llmtospeech, an architecture to synthesize speech while text is being generated by an LLM which yields significant latency reduction.
\llmtospeech\ mimics the predictions of a non-streaming teacher model while limiting the exposure to future context in order to enable streaming. 
It exploits the hidden embeddings of the LLM, a by-product of the text generation that contains informative semantic context.
Experimental results show that \llmtospeech\ maintains the teacher's quality while reducing the latency to enable natural conversations.    
\end{abstract}

\begin{keywords}
Incremental TTS, Speech Generation, Large Language Models (LLMs)
\end{keywords}

\section{Introduction}
\label{sec:intro}
The appearance of conversational Large Language Models (LLMs)~\cite{chatgpt, bard} 
has revolutionized the scope of interactions with computers.
By leveraging the principles of self-learning and vast amounts of unlabeled training data, LLMs have established a new state-of-the-art across a variety of tasks, and showed great promise as a tool to augment human intelligence.
Currently, the interaction with LLMs is usually facilitated through text even though in many applications, 
such as driving assistance, the spoken modality is far more preferable, intuitive, and safe.
As a result, pure audio-based LLMs~\cite{borsos2023audiolm, kreuk2023audiogen} are gaining interest in the community, though their semantic language understanding capabilities still lag behind textual LLMs. 
A simple alternative approach for spoken LLMs that would address this issue would be to couple a text-based language model with a neural Text-To-Speech (TTS) system capable of producing high-fidelity speech samples \cite{shen2018natural, wang2023neural}. TTS models, however, often require an entire sentence to generate natural speech, resulting in notable \emph{latency} when combined with an LLM that typically generates text in a slow autoregressive fashion. This work tackles several challenges that arise when trying to read aloud text generated by an LLM, incrementally and with minimal delay.

TTS systems typically adopt a two-step process, first converting graphemes to phonemes (G2P) and then converting phones to speech, as this process often achieves improved quality and stability over character-to-speech approaches for languages with irregular orthography (e.g. English, French)~\cite{Fong-Taylor:19,Taylor-Richmond:19}.
Context-dependent G2P methods look beyond the unigram level to improve the quality of the phonetic prediction~\cite{ploujnikov2022soundchoice-interspeech,rezackova21_interspeech,zhu2022byt5-interspeech}. 
Such models can, e.g., better account for cross-word-boundary flapping, vocalic reduction, and heteronym disambiguation.
In these models, however, the context required for disambiguation may be long, rendering them unsuitable for streaming.
Incremental TTS works create low-latency TTS systems with limited lookahead (i.e. exposure to future context) and minimal degradation~\cite{ma-etal-2020-incremental,chen2021speech,wu21b_interspeech,ellinas2020high-interspeech}. In most scenarios, the entire text is available before synthesis, and the focus is on reducing algorithmic delay.
These methods may, e.g., run a full-sentence lightweight G2P module before synthesis without significantly contributing to the overall delay.
When considering a slow incoming stream of generated text, though, this assumption no longer holds.

This paper introduces \llmtospeech, a system that integrates a generative LLM with a streamable TTS system. 
\llmtospeech\ can speak the text aloud while it is being generated by the LLM, without compromising correctness or naturalness.
\llmtospeech\ utilizes LLM embeddings, a \emph{by-product} of the text generation that contains semantic information and might compensate for the lack of future context in streaming. \llmtospeech\ consists of three parts (Fig.~\ref{fig:diagram}):
\begin{figure}[htb]
\vspace{-2mm}
\centering
\centerline{\includegraphics[width=\linewidth]{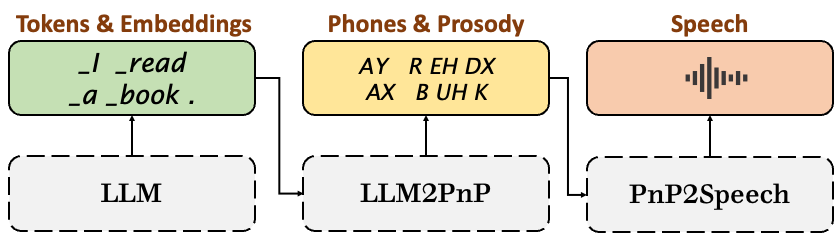}}
\setlength{\abovecaptionskip}{8pt}
\setlength{\belowcaptionskip}{0pt}
\caption{Streaming speech synthesis during text generation. Word tokens and embeddings are incrementally generated by the LLM, sent to \llmtophones\ to produce phones and prosody (\emph{PnP}), which are sent to \phonetospeech\ to produce audio.}
\vspace{-2mm}
\label{fig:diagram}
\end{figure}
\begin{enumerate}[wide,topsep=1pt,itemsep=0ex,partopsep=1ex,parsep=1ex]
    \item A pretrained LLM, which we deliberately freeze due to the vast computational and human effort invested in bringing it to its final state \cite{NEURIPS2022_b1efde53}.
    \item \llmtophones:  
    an adaptor converting the LLM outputs to Phones and Prosody (\emph{PnP}), which is described in Sec.~\ref{sec:llmtophones}.
    \item \phonetospeech: a streamable version of the TTS system in~\cite{Shechtman-Fernandez:23}, which operates on chunks of \emph{PnP} (see Sec.~\ref{sec:p2s}).
\end{enumerate}
\llmtophones\ is trained on a large textual dataset via offline-to-streaming knowledge distillation \cite{povey2018time,kurata2020knowledge} during which it attempts to mimic the predictions of a teacher model that has access to the full text. 
We evaluate \llmtospeech\ against the offline teacher TTS for phonetic accuracy as well as the quality of the synthesized speech. We demonstrate that the overall quality is maintained using both objective and qualitative measures.
\llmtospeech\ shows impressive prosodic predictions even for expressive inputs (e.g. happy, empathetic, uncertain), and can also synthesize interjections and filled pauses (e.g., {\em hmm}, {\em uh-huh}, {\em oh}, etc.).
\footnote{\href{https://ibm.biz/BdMe5X}{Audio samples can be found here: https://ibm.biz/BdMe5X}}

The work proposed here makes the following novel contributions to the field of conversational speech synthesis:
\begin{enumerate}[wide,topsep=1pt,itemsep=0ex,partopsep=1ex,parsep=1ex]
    \item It introduces a pipeline that converts an LLM output text into expressive speech incrementally and with low delay.
    \item It proposes a streaming knowledge-distillation method for training \emph{PnP} models based on large textual datasets.
    \item It quantifies the contributions of LLM hidden embeddings to the task of \emph{PnP} prediction.
\end{enumerate}

\section{Method}
\label{sec:method}

\subsection{Dataset creation}
\label{sec:dataset_creation}

\begin{figure}[htb]
\vspace{-2mm}
\centering
\centerline{\includegraphics[width=\linewidth]{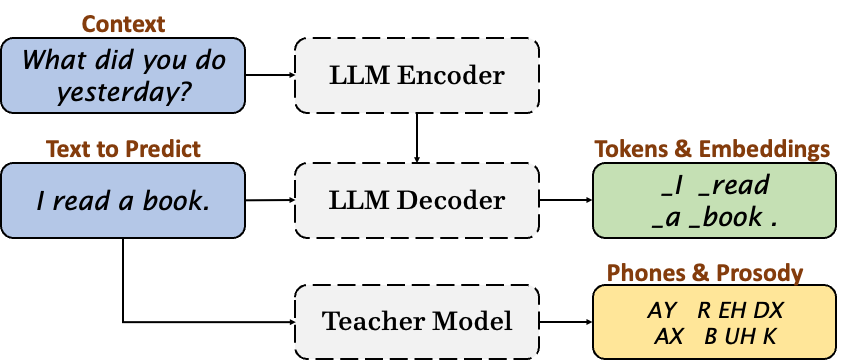}}
\setlength{\abovecaptionskip}{6pt}
\setlength{\belowcaptionskip}{-10pt}
\caption{Dataset creation. Above: extracting tokens and contextual embeddings from the LLM, which are conditioned on the \emph{context}. 
Below: pseudo-labeling for phones and prosody, generated by the teacher model.}
\label{fig:data_gen}
\end{figure}

As \llmtospeech\ uses the tokens and embeddings of an LLM, it is trained for a specific LLM.
We experiment with the \texttt{T5} language model \cite{raffel2020exploring-jmlr}, due to its capabilities to perform diverse conditional generation tasks. 
Specifically, we use \texttt{T5-lm-adapt}, which was finetuned for text completion. 
We construct our training dataset based on the C4 (Common Crawl Cleaned Corpus) dataset~\cite{raffel2020exploring-jmlr}, which was also used to train \texttt{T5}. 
As C4 contains 365M samples, we consider only a random fraction of the dataset containing 3M training and 130K validation samples. 
Each sample in C4 contains a paragraph, which we split randomly
into two parts: \emph{context} and \emph{text-to-predict (t2pred)},
such that \emph{t2pred} has 1-5 sentences. 
We simulate conditional text generation where the LLM is prompted with a \emph{context} and generates \emph{t2pred}
by inputting \emph{context} to the \texttt{T5} encoder and \emph{t2pred} to the \texttt{T5} decoder.\footnote{When using decoder only LLMs, the text split is not needed.} 
The inputs to the \llmtophones\ training task are the word tokens for \emph{t2pred} and their contextual embeddings (see Fig.~\ref{fig:data_gen}). 
We also obtain the \emph{PnP} annotations for \emph{t2pred} using a teacher model with no lookahead restrictions.
The \emph{PnP} teacher model 
is a rules-based G2P model predicting the phonetic sequence and phrase type, followed by a neural model predicting Hierarchical Prosodic Controls (HPCs)~\cite{Shechtman-Fernandez:21} for an expressive conversational speaking style~\cite{Fernandez-Shechtman:22} and phone durations. 
HPCs are speaker-agnostic prosodic statistics that can be calculated from recordings at various resolutions and have been used 
for various tasks~\cite{Fernandez-Shechtman:22, Shechtman-Fernandez:23}. 
We use duration and pitch HPCs~\cite{Shechtman-Fernandez:23} augmented with maximal log-energy, evaluated at the sentence, word, and phone hierarchies. 
To account for text-normalization expansions (e.g. converting \emph{23} to \emph{twenty-three}), we differentiate between \emph{regular} and \emph{inner} word separators, where inner word separators are placed only when expansions occur. During inference, \llmtophones\ synthesizes a word until reaching a regular word separator, and then waits for the LLM to generate the next word.

\subsection{\llmtophones{}}
\label{sec:llmtophones}
The \llmtophones\ is a transformer encoder-decoder model, augmented with attention restrictions. 
The encoder input is a sequence of tokens (word pieces) and their contextual embeddings, obtained from the hidden layers of the LLM, which are projected to the encoder using a linear layer.
The decoder outputs are used by three prediction modules that predict the identity, prosodic features, and phrase type of the next phone.

\subsubsection{Restricted attention}
\begin{figure}[htb]
\vspace{0mm}
\centering
\begin{subfigure}{0.25\textwidth}
 \centering
 \includegraphics[width=\textwidth]{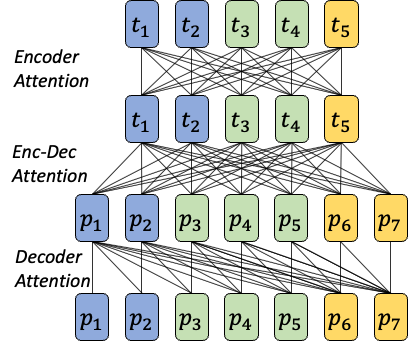}
 \caption{Regular attention}
 \label{fig:regular_attention}
\end{subfigure}
\hfill
\begin{subfigure}{0.212\textwidth}
 \centering
 \includegraphics[width=\textwidth]{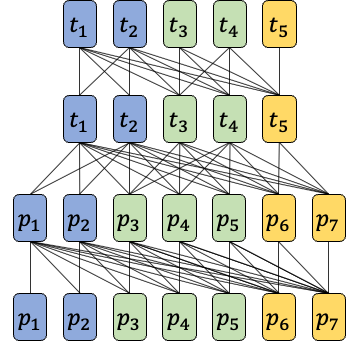}
 \caption{Restricted attention}
 \label{fig:restricted_attention}
\end{subfigure}
\hfill
\setlength{\abovecaptionskip}{4pt}
\setlength{\belowcaptionskip}{-10pt}
\caption{Regular/restricted attention (colored by word). In (a), every encoder/decoder token can attend to every encoder token.
In (b), dependence on future context is limited to the current word ($L=0$) as in Eqs.~\ref{eq:encoder_attn}--\ref{eq:decoder_attn}, allowing streaming.}
\vspace{-3mm}
\label{fig:attention_types}
\end{figure}

To restrict dependence on future context, we formalize restricted attention with a fixed word lookahead $L$. 
We first define the sequence of words $w_1,...,w_n$,
word tokens  $t_1,...,t_m$ and
\emph{PnP} tokens $p_1,...,p_k$.
Each word token $t_j$ is a part of some word $w_i$, which we denote by $\word(t_j)=i$. 
Similarly, for every phoneme $p_j$ and its word $w_i$, we denote $\word(p_j)=i$.
We now denote that an output token $y$ 
can attend to an input token $x$ by $y\to x$.
In regular encoder attention and encoder-decoder attention \cite{vaswani2017attention}, $\forall i,j$ we have  
$t_i\to t_j$ and $p_i\to t_j$ (see Fig~\ref{fig:regular_attention}). 
We define restricted encoder and encoder-decoder attention as follows:
\vspace{-0.135cm}
\begin{equation}
t_i \to t_j \iff \word(t_j) \le \word(t_i)
\label{eq:encoder_attn}
\vspace{-0.135cm}
\end{equation}
\begin{equation}
p_i \to t_j \iff \word(t_j) \le \word(p_i) + L    
\label{eq:decoder_attn}
\end{equation}
We chose to use $L$ in the encoder-decoder attention since it would not grow in consecutive decoder layers, unlike encoder attention, where the lookahead would grow linearly with the number of layers.
Fig.~\ref{fig:restricted_attention} visualizes restricted attention, highlighting that $t_1 \not \to t_3$ and $p_1 \not \to t_3$, which
ensures the prediction of $p_1$ would not depend on $t_3$. 


\subsection{\phonetospeech{}}
\label{sec:p2s}

\begin{figure}[htb]
\centering
\vspace{0mm}
\centering
\begin{subfigure}{0.25\textwidth}
 \centering
 \includegraphics[width=\textwidth]{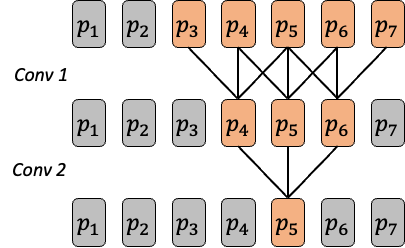}
 \caption{CNN}
 \label{fig:regular_conv}
\end{subfigure}
\hfill
\begin{subfigure}{0.212\textwidth}
 \centering
 \includegraphics[width=\textwidth]{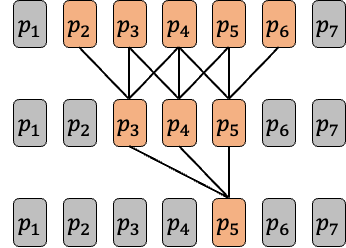}
 \caption{LC-CNN ($la=1$)}
 \label{fig:restricted_conv}
\end{subfigure}
\hfill
\setlength{\abovecaptionskip}{4pt}
\setlength{\belowcaptionskip}{-1pt}
\caption{Visualising the receptive field of $p_5$ (in orange) at the output of two stacked convolution layers with a $k=3$ kernel.
In Fig~\ref{fig:regular_conv}, the right receptive field (i.e. lookahead) grows with the number of layers.
In Fig.~\ref{fig:restricted_conv}, the 2nd convolution is skewed so that the final lookahead would be exactly $la=1$.}
\label{fig:conv_types}
\end{figure}

\phonetospeech\ is a streamable version of the HPC-based Parallel Prosody Transfer (PPT) model~\cite{Shechtman-Fernandez:23}, composed of an acoustic model, based on the non-attentive Tacotron (NAT) backbone~\cite{Shen-Jia:20}, followed by a lightweight and streamable LPCNet vocoder~\cite{valin2019lpcnet}.
\phonetospeech\ 
operates on small input chunks instead of the entire sequence, 
and requires the changes to~\cite{Shechtman-Fernandez:23} described below. 
First, BLSTMs~\cite{graves2005framewise} were chunked, thus becoming LC-BLSTM~\cite{zhang2016highway} layers with zero lookahead ($la=0$).
Next, the growing right-receptive field of convolution neural network (CNN) layers was addressed using \emph{lookahead-constrained CNNs} (LC-CNNs). 
Symmetric-kernel convolutions are used as long as the lookahead constraint is met; otherwise, skewed-kernel convolutions (a generalization of causal convolutions~\cite{oord2016wavenet})
are applied, resulting in a constrained lookahead (See Fig.~\ref{fig:conv_types}).
Finally, on inference, we include guardbands when chunking the \emph{PnP}-to-frame Gaussian upsampling matrix~\cite{Shen-Jia:20}, as the upsampling depends on the adjacent future \emph{PnP}. 



We train \phonetospeech\ on a 6.5-hr, proprietary, conversational speech corpus recorded by a professional US-English female speaker. This set
contains a variety of expressive dialog acts and interjections and has been described in~\cite{Fernandez-Shechtman:22}. 
%
To further improve the performance of interjections and expressive styles, we finetuned \llmtophones\ on the same corpus described above, training it to predict the conversational set of \emph{PnP}s from the conversational text.




\section{Experiments}
\label{sec:experiments}
In the following experiments, \llmtophones\ has 4 encoder and 6 decoder layers, each one with token/feedforward dimensions of 512/768 and 4 attention heads. 
\llmtophones\ has a single word lookahead and uses the \texttt{T5-Base} embeddings from layers 2, 6, and 10 (out of 12 layers, numbered from input to output).

\phonetospeech\ has a frame size of $256$ samples for $22kHz$-sampled speech. 
Its phonetic encoder has 3 LC-CNN layers and a single chunked BLSTM layer, followed by Gaussian upsampling, then the autoregressive LSTM decoder, and the 5-layered LC-CNN PostNet.
LC-CNN layers have a kernel size of $5$ and a lookahead of 2. 
Chunked BLSTM layers have a chunk size of 4, and the Gaussian upsampler uses a 2-phone guardband.
The proposed \phonetospeech\ system results in an algorithmic delay of $6$ \emph{PnP} tokens plus 2 frames, which is approximately equivalent to one word (where word separators and pauses are also considered as \emph{PnP} tokens). 
Consequently, the total \llmtospeech\ lookahead sums up to 2 words.



\subsection{Quality and naturalness assessment}
\label{sec:mos}
We crowd-sourced a Mean Opinion Score (MOS) listening test to evaluate \llmtospeech\ quality and naturalness, comparing three systems:
\begin{enumerate}[wide,topsep=1pt,itemsep=0ex,partopsep=1ex,parsep=1ex]
\item Teacher: The non-streamable, teacher \emph{PnP} model, followed by a non streaming TTS \cite{Shechtman-Fernandez:23}.
\item \llmtospeech\ (Ours): using \llmtophones\ to extract \emph{PnP}, followed by \phonetospeech.
\item Stream-Teacher: forcing the teacher into the same lookahead as \llmtospeech, by using the teacher G2P on text prefixes with 1-word lookahead, the prosody model with lookahead restrictions, and \phonetospeech\ for synthesis.
\end{enumerate}
\begin{table}[htb]
\centering
\begin{tabular}{ccccc}
\toprule
Model & Lookahead           &   MOS\\
\midrule
Teacher& $\infty$    &   $4.10\pm 0.04$\\
LLM2Speech & 2        &  \textbf{4.12} $\boldsymbol{\pm}$ \textbf{0.04}\\
Stream-Teacher & 2            &   $3.46\pm 0.06$\\
\bottomrule
\end{tabular}
\setlength{\abovecaptionskip}{4pt}
\setlength{\belowcaptionskip}{-5pt}
\caption{Listening test results, reporting the MOS and the 95\% confidence interval. 
}
\vspace{-2mm}
\label{table:mos_main}
\end{table}

The synthesized audio was evaluated on 45 conversational texts, and rated for overall quality and naturalness by 25 native listeners on a standard MOS 5-point scale.  
Results in Table~\ref{table:mos_main} show no statistically significant difference between the teacher and the streamable \llmtospeech.
Note that the teacher makes predictions based on the entire text (which is unrealistic for streaming) and
utilizes 
the sub-style labeling of the text (e.g., empathetic, happy, etc.) that \llmtospeech\ is not exposed to.

\subsection{G2P ablation study}
\label{sec:g2p_performance}
We evaluate the G2P performance of \llmtophones\ by measuring the word error rate (WER) on the C4 validation set. 
We accentuate the differences by presenting results on the following challenging subsets:
\begin{inparaenum}[(i)]
    \item \emph{Rare}: least common words covering 20\% of the text,
    \item \emph{Norm}: words expanded by normalization, e.g. 23, and
    \item \emph{OOV}: words unseen during training.
\end{inparaenum} 

Incremental TTS methods trade off latency and performance, which are determined by the lookahead. 
In Table~\ref{table:lookahead_size}, we show the effect on G2P performance by modifying the lookahead of \llmtophones.  
Results suggest the first lookahead word is quite significant while the second yields smaller benefits.
This observation may be partly explained by post-lexical processes in US English which influence the pronunciation of a word depending on the word that follows.
\begin{table}[htb]
\centering
\begin{tabular}{ccccc}
\toprule
Lookahead & All & Rare & Norm & OOV \\
\midrule
0 & 6.40 & 14.99 & 21.71 & 37.33\\
1 & 1.95 & \hphantom{0}2.71 & \hphantom{0}6.28 & 18.90 \\
2 & 1.69 & \hphantom{0}2.55 & \hphantom{0}6.02 & 18.62\\
$\infty$    & \textbf{1.31} & \hphantom{0}\textbf{2.17} & \hphantom{0}\textbf{5.28} & \textbf{17.28}\\
\bottomrule
\end{tabular}
\setlength{\belowcaptionskip}{-5pt}
\caption{Lookahead influence on G2P performance, measured by Word Error Rate (\%), on all words and challenging subsets as defined in Sec~\ref{sec:g2p_performance}.
}
\vspace{-2mm}
\label{table:lookahead_size}
\end{table}

In the experiments described above we made use of the \texttt{T5-Base} model.
However, larger language models are more commonly used due to their improved performance, and the issue of  
using more or fewer LLM hidden layers could also be considered.
We investigate the influence of the LLM embeddings used by \llmtophones\ on the G2P performance by adding and removing embedding layers from \texttt{T5-Base}, and by also making use of the 24-layer \texttt{T5-Large} and \texttt{T5-XL} models.
Table~\ref{table:embedding} suggests that both adding more layers and increasing the LLM size improve the G2P performance.
However, the benefits gained by the choice of LLM embeddings are smaller than those gained by an additional word lookahead.
\begin{table}[htb]
\centering
\begin{tabular}{cccccc}
\toprule
LLM & Emb Layers & All & Rare & Norm & OOV \\
\midrule
-& -             & 2.10 & 2.93 & 6.61 & 20.05 \\
\midrule
Base & $6$      & 1.98 & 2.78 & 6.51 & 19.84 \\
 & $2,6,10$ & 1.95 & 2.71 & 6.28 & 18.90 \\
 & $2,4,6,8,10$        & 1.93 & 2.69 & 6.21 & 18.66 \\
\midrule
Large & $6,12,18$ &  1.94 & 2.71 & 6.31 & 18.91\\
\midrule
XL & $6,12,18$ &   \textbf{1.89} & \textbf{2.62} & \textbf{6.02} & \textbf{18.31} \\

\bottomrule
\end{tabular}
\setlength{\belowcaptionskip}{-5pt}
\caption{Embedding influence on G2P as in Table~\ref{table:lookahead_size}.}
\vspace{-2mm}
\label{table:embedding}
\end{table}

\subsection{Prosody ablation study}
\label{sec:prosody_ablation}
To estimate prosodic quality, we conducted ABX preference tests, where we compared \llmtospeech\ synthesized audio (A) with another system's audio (B) on the same texts as in Sec.~\ref{sec:mos}.
Each pair of audio samples was rated by 25 distinct listeners, who were asked to rate their preference on a scale of $[-2, -1, 0, 1, 2]$, where -2 is ``strongly prefer A'' 2 is ``strongly prefer B'' and 0 is ``no preference'' 
We compared to a variant of \llmtospeech\:
\begin{inparaenum}[(i)]
    \item without finetuning on the conversational corpus (NoFT), 
    \item without LLM embeddings (NoEmb),
    \item with \texttt{T5-XL} embeddings (T5XL), and
    \item with a lookahead of $L=2$ for \llmtophones\ (LA2).
\end{inparaenum}
For fairness, we removed interjections from the texts in (i), as NoFT was not exposed to them during training.
Results in Table~\ref{table:ablation_listening_test} suggest that finetuning improved the overall naturalness, yet other changes did not yield a significant difference.

\begin{table}[htb]
\centering
\begin{tabular}{ccccccccc}
\toprule
\multirow{2}{*}{Method B}& \multicolumn{5}{c}{Vote Distribution (\%)} & \multirow{2}{*}{Avg Score}\\ \cmidrule(l){2-6}
 & -2 & -1 & 0& 1 & 2&  \\ 
\midrule
\hphantom{O}NoFT & 8.9 & 30.1& 30.2& 24.6 & 6.2 & \textbf{-0.110} \\
NoEmb & 5.8 & 26.8 & 35.8 & 25.3 & 6.3 & -0.005 \\
\hphantom{O}T5XL & 5.0 & 25.9 & 36.7& 28.0 & 4.3 & \hphantom{-}0.007\\
\hphantom{Oo}LA2 & 8.0 & 30.8 & 25.9 & 27.8 & 7.4 & -0.042\\
\bottomrule
\end{tabular}
\setlength{\belowcaptionskip}{-10pt}
\caption{ABX preference test results comparing \llmtospeech\ (A) to another system (B).
Negative scores mean A is preferred, and results indicating significant differences ($p<0.01$) are bolded.}
\vspace{-4mm}
\label{table:ablation_listening_test}
\end{table}

\section{Discussion}
\label{sec:discussion}
Motivated by spoken conversational AI, we investigated reading aloud LLM-generated text with low latency, paving the way for natural AI conversations.
We described simple mechanisms to limit the lookahead in attention and convolution layers, with which we build a low-latency conversational TTS system.
We found that streaming TTS benefits from offline-to-streaming distillation using large textual datasets, even when the texts lack a conversational style.
Moreover, the LLM embeddings improved the phonetic prediction, yet did not yield a significant improvement in the prosodic quality.


In future work, we aim to improve the audio quality by utilizing natural speech and by extracting additional cues from the LLM  such as emotions. 
More broadly, we intend to create a low-latency spoken dialogue system, powered by an LLM semantic backbone. The system would consist of a speech recognition model, followed by a conversational LLM, which is coupled with \llmtospeech\ to produce speech incrementally, with low latency.


\vfill\pagebreak


\bibliographystyle{IEEEbib}
\bibliography{main}
\end{document}